\documentclass[amsfonts,amssymb,aps,twocolumn]{revtex4}

\usepackage{epsfig}

\def\beq{\begin{equation}}
\def\eeq{\end{equation}}
\def\2{\mbox{$1\over2$}}
\def\6{\langle}
\def\9{\rangle}

\newcounter{step} 
\newcommand{\debprotocol}[1]{
\begin{center}{\bf #1}
\end{center}
\begin{list}{}{
\setlength{\leftmargin}{0pt}
}}

\newcommand{\finphase}{\end{list} \mbox{}}
\newcommand{\finprotocol}{\end{list} \setcounter{step}{0}}

\begin{document}
\title{Super Selection Rules in Quantum Cryptography}

\author{Dominic Mayers}
\address{Universit\'e de Sherbrooke, Qu\'ebec, Canada}

\begin{abstract}
It is believed that superselection rules in quantum mechanics can
restrict the possible operation on a qbit. If this was true, the model
used by Mayers for the impossibility of bit commitment and by Kitaev
for the impossibility of coin flipping would be inadequate. We explain
why this is not the case.  We show that a charge superselection rule
implies no restriction on the operations that can be executed on any
individual qbit.
\end{abstract}
\maketitle 

It is well known that superselection rules restrict the allowed
operations in quantum mechanics.  Therefore, it is natural to suggest
\cite{popescu02,personal02} that the models used by Mayers to prove
the impossibility of bit commitment \cite{mayers97} and by Kitaev for
the impossibility of coin flipping \cite{kitaev02} are inadequate.
Here, we show that to the contrary the models originally used by
Kitaev and Mayers for their respective impossibility proofs are valid
even in the context of superselection rules.

Superselection rules are associated with conserved quantities such as
the number of fermions.  Let $H_F$ be the span of $|0\rangle^F$ and
$|1\rangle^F$, a $0$ fermion state and a $1$ fermion state,
respectively.  Let $H_B$ be the polarisation state space of a
photon. It may appear impossible to execute the swap gate $S$ on $H_F
\otimes H_B$ which executes $|x\rangle^F 
|y \rangle^B \leftrightarrow
|y\rangle^F |x \rangle^B$ because the fermion number is not preserved
when $x \neq y$.  The basic idea to unvalidate this argument can be
understood with an analogy with a common phenomena which also appears
to violate a conservation law.  Consider a photon that is reflected by
a mirror. The dynamic of the photon, if considered separately, is
essentially a unitary transformation which violates the preservation
of momemtum. Of course, momentum is preserved because the mirror
absorbs the difference in momemtum.  Yet the mirror does not get
entangled with the photon.  The mirror can be considered as a
catalyser for a unitary transformation that would otherwise be
impossible because it violates a conservation law.  In a similar way,
we will show that by using extra degrees of freedom as a catalyser, we
can execute the swap operation on $H_F \otimes H_B$, an operation that
would otherwise be impossible.  This swap operation share a similarity
with the reflection of a photon on a mirror.  However, this swap
operation is more powerful because, as we will see, it implies that
all operations on $H_F \otimes H_E$, where $H_E$ is any extra system,
are allowed despite superselection rules.  

There is a subtle point here. The state of these extra degrees of
freedom must itself be a superposition of states with different
fermion numbers.  The existence of such initial states in nature would
not violate any conversation law by itself. For example, even tough
momemtum is a conserved quantity, a superposition of states with
different momemtum is a common phenomena.  Nevertheless, one might ask
if such states really exist in the case of all conservation laws.  It
turns out that, if these initial states are not allowed, the original
proof that bit commitment is impossible can easily be generalized.
Therefore, in both cases, the conclusion is essentially the same.  In
this paper, we focus on the case where initial states of the form
$(1/\sqrt 2)(|0\rangle^F + |1\rangle^F)$ are allowed, but the easier
case where such a superposition is not allowed will be considered
also.

There is a similarity between the constraint imposed by the fermion
number conservation superselection rule and the one imposed by the use
of classical information in quantum protocols. In both cases, it was
believed for some times that the constraint could perhaps be used to
build a bit commitment protocol. In both cases, it turned out to be
useless.  Before the general impossibility result, we knew that bit
commitment was impossible if we accept the model where classical
information can be manipulated as if it was quantum information
\cite{trouble,lc97}.  Of course, researchers seriously looked for a
bit commitment protocol that makes use of the additional constraint
associated with classical information. The proof that this additional
constraint cannot help is not so complicated but yet it took six
months and many unsuccessful attempts before the author realised that
classical information was actually useless \cite{mayers97}.  In the
case of the fermion number conservation superselection rule, a few
researchers also attempted to make use of the corresponding additional
constraint \cite{personal02}, but it took only a few weeks in this
case to realize that the additional constraint is useless, and this is
the subject of the current paper.

To our knowledge, this is the first analysis of superselection rules
in quantum cryptography.  It is interesting to see that our analysis
is inspired by a very simple phenomena, the reflexion of a photon in a
mirror.  We will not refer to this simple phenomena anymore, but
anyone who knows why the mirror is not entangled with the photon
should appreciate some connection.  In the first section, we describe
the model which includes superselection rules.  In the second section,
we show that, if arbitrarily small errors are ignored, this model
implies no restriction on the possible operations on a given qbit.

\subsection*{The Model} \label{introduction_sect}
We assume that the Hilbert space $H^{(all)}$ for the protocol is the
span of a set of orthogonal states
\[
|(q_x)_{x \in {\cal M}} \rangle^{(all)} =
\otimes_{x \in {\cal M}} |q_x\rangle^{(x)},
\]
where ${\cal M}$ is the set of possible modes and, for every $x \in
{\cal M}$, $q_x \in \{0,1 ,2 ,\ldots, q_x^{max} \}$ where $q_x^{max} =
1$ for fermion modes and $q_x^{max}$ depends on the protocol for boson
modes.  The state space $H^{(all)}$ is simply the ordinary tensor
product of the state spaces $H_{x}$ associated with the modes $x \in
{\cal M}$.  A superselection rule does not change the state space
$H^{(all)}$, but restricts the possible operations on this space.

{\em Superselection rules.}  A superselection rule $R$ requires that,
for a subset ${\cal M}_R \subseteq {\cal M}$, the quantity $Q_R =
\sum_{x \in {\cal M}_R} q_x$ is preserved by every allowed
transformation.  More precisely, the allowed unitary transformations
are block diagonal and each block is associated with a given value for
$Q_R$.  We assume that every participant has access to a countably
infinite number of extra modes $x \in {\cal M}_R$ which are normally
not used in the protocol (i.e., normally, $q_{x} = 0$ for these
modes).  For example, these extra modes can lie in some unused
locations on the participant's side.

{\em Local transformations.} We assume that ${\cal M}$ is partitioned
between Alice and Bob in two subsets of modes ${\cal M}^A$ and ${\cal
M}^B$.  We denote $H^A$ the span of the orthogonal states
\[
|(q_x)_{x \in {\cal M_A}} \rangle^{(A)} =
\otimes_{x \in {\cal M_A}} |q_x\rangle^{(x)}.
\]
Similarly, we denote $H^B$ the span of the orthogonal states
\[
|(q_x)_{x \in {\cal M_B}} \rangle^{(B)} =
\otimes_{x \in {\cal M_B}} |q_x\rangle^{(x)}.
\]
We have $H^{all} = H^A \otimes H^B$.  Locally, Alice can
only execute unitary transformations of the form $U^A \otimes \bf
I$ which moreover respect the superselection rules.  Locally, Bob can
only execute unitary transformations of the form $\bf I \otimes
U^B$ which moreover respect the superselection rules.

{\em Measurements.}  Alice has access to a countably infinite set of
``free'' modes ${\cal M}^A_F \subseteq {\cal M}^A - {\cal
M}_R$. Similarly, Bob has access to a countably infinite set of
``free'' modes ${\cal M}^B_F \subseteq {\cal M}^B - {\cal M}_R$. We
denote ${\cal M}_F = {\cal M}^A_F \cup {\cal M}^B_F$.  Without loss of
generality, we assume that only the Hilbert spaces $H^A_F = \otimes_{x
{\cal M^A_F}} H_x$ and $H^B_F = \otimes_{x {\cal M^B_F}} H_x$ can be
measured by Alice and Bob, respectively. However, there is no
restriction on these measurements.

{\em Communication.} Alice communicates information to Bob by
transferring control over a mode $x \in {\cal M}^A$ to Bob.  Before
the communication, we have $x \in {\cal M}^A$.  After the
communication, we have $x \in {\cal M}^B$.  Note that the sets ${\cal
M}^A$ and ${\cal M}^B$ can change during the execution of the
protocol, but the sets ${\cal M}_R$ and ${\cal M}_F$ are fixed.

\subsection*{Why there is no restriction on the possible operations}
It may appear that a superselection rule $R$ restricts the possible
operations on every state space $H_x$ associated with a mode $x \in
{\cal M}_R$.  However, in fact, the restriction only applies to the
tensor product $\otimes_{x \in {\cal M}_R} H_{x}$, not to each state
space $H_{x}$ individually. We will show that, if a participant has
access to as many extra unused modes in ${\cal M}_R$ as needed, the
superselection rule $R$ imposes essentially no restriction on the
operations that can be executed by this participant on a given state
space $H_{x}$ with $x \in {\cal M}_R$.  We will only do the case
where $H_{x}$ is a qbit, but the generalisation to higher dimensions
is not difficult.

Consider the most general qbit state $\alpha |0\rangle^{(x)} + \beta
|1\rangle^{(x)}$ for a mode $x$. We will show how the participant can
swap this state with the state of a free mode $z$ initially in the
state $|0\rangle^{(z)}$.  The swapping will only be approximative, but
it will be arbitrarily close to perfect. The participant uses $n-1$
extra modes $\vec y = y_1,\ldots, y_{n-1}$ and prepares the state
$(1/\sqrt n) \sum_{j = 0}^{n-1} |j\rangle^{(\vec y)}$ where
$|j\rangle^{(\vec y)} = |1,\ldots, 1, 0,\ldots, 0\rangle^{(\vec y)}$
contains exactly $j$ modes $y_1, \ldots y_j$ with $q_{y_j} = 1$. The
overall state is
\begin{eqnarray*}
(1/\sqrt{n}) (\alpha |0 \rangle^{(x)} + \beta |1\rangle^{(x)}) \otimes
(\sum_{j = 0}^{n-1} |j \rangle^{(\vec y)} ) \otimes |0\rangle^{(z)}
\end{eqnarray*}
which can be rewritten as
\begin{eqnarray*}
\lefteqn{(1/\sqrt n)( \alpha |0,0,0\rangle^{(x, \vec y,z)} + 
\beta |1, n-1, 0\rangle^{(x, \vec y, z)}} \\
& & + 
\sum_{j = 1}^{n-1} 
\alpha |0, j, 0\rangle^{(x, \vec y, z)} 
+ \beta |1, j - 1, 0\rangle^{(x, \vec y, z)} ).       
\end{eqnarray*}
Now, each term $\alpha |0, j, 0 \rangle^{(x, \vec y, z)} + \beta |1, j
- 1, 0\rangle^{(x, \vec y, z)}$ in the sum have $Q_R = j$. Any
operation that preserves $Q_R$ is allowed. In particular, the
operation $S$ on $H_{x} \otimes H_{\vec y}$ that executes the
swapping $|0, j \rangle^{(x, \vec y)} \leftrightarrow |1, j -
1\rangle^{(x, \vec y)}$, for $j = 1, \ldots, n-1$, is allowed.  For $1
\leq Q_R \leq n-1$, the participant can execute the following. First,
he executes a CNOT with the mode $x$ as the source and the free mode
$z$ as the target.  For $Q_R = j$, we obtain the component
\[
\alpha |0, j, 0 \rangle^{(x, \vec y, z)} + \beta |1, j - 1,
1\rangle^{(x, \vec y, z)} ).
\]
Next, conditioned on the free mode $z$, he executes the swapping
$S$. For $Q_R = j$, we obtain the component
\begin{eqnarray*}
\lefteqn{ \alpha |0, j, 0 \rangle^{(x, \vec y, z)} + \beta |0, j,
1\rangle^{(x, \vec y, z)} )} \\ && \quad = |0, j \rangle^{(x, \vec y)}
\otimes (\alpha |0 \rangle^{(z)} + \beta |1\rangle^{(z)}).
\end{eqnarray*}
If we sum all the components, the resulting state is
\begin{eqnarray*}
\lefteqn{(1/\sqrt n)( \alpha |0,0,0\rangle^{(x, \vec y, z)} + 
\beta |1, n-1, 0\rangle^{(x, \vec y, z)}} \\
& & + |0\rangle^{(x)} \otimes 
(\sum_{j = 1}^{n-1} |j \rangle^{(\vec y)}) \otimes 
(\alpha |0
\rangle^{(z)} + \beta |1\rangle^{(z)}).       
\end{eqnarray*}
Note that we have $\Pr(Q_R = 0) = \alpha^2/n$, $\Pr(Q_R = n) =
\beta^2/n$ and $\Pr(Q_R = j) = 1/n$, for $j = 1, \ldots, n-1$, as
expected.  No conservation law is violated.  Fortunately, this state
is arbitrarily close to the state
\begin{eqnarray*}
(1/\sqrt{n-1}) |0\rangle^{(x)} \otimes (\sum_{j = 1}^{n-1} |j
\rangle^{(\vec y)} ) \otimes (\alpha |0 \rangle^{(z)} + \beta
|1\rangle^{(z)})
\end{eqnarray*}
which corresponds to a swapping of the two modes $x$ and $z$. This
state corresponds to the distribution of probability $\Pr(Q_R = j) =
1/(n-1)$, $j = 1, \ldots, n-1$, which is arbitrarily close to the
actual distribution of probability associated with the true final
state. So, the participant has essentially executed a swap operation
$S_{x \leftrightarrow z}$ between the two modes $x$ and $z$, despite
the fact that $z$ is a free mode and $x$ is apparently restricted by a
superselection rule.

Let $H_E$ be the state space of some extra system. It is not difficult
to see that, given such a swap operation, one can execute the most
general transformation $U$ on $H_{x} \otimes H_E$.  First, one execute
the swap operation $S_{x \leftrightarrow z}$ on $H_{x} \otimes
H_{z}$. Second, one executes $U$ on $H_{z} \otimes H_E$ in the same
way it would have been executed on $H_{x} \otimes H_E$.  Finally, one
execute the inverse swap operation $S^{\dagger}_{x \leftrightarrow z}$
on $H_{x} \otimes H_{z}$.

\subsection*{The easier case}
Now, we consider the easier case in which an initial superposition of
states with different values for the conserved quantity is not
allowed.  Note that, on Alice's side, a mixture of different values
for $Q^A_R$ can be purified with a state that has a fixed value for
$Q^A_R$, the maximum of the possible values for $Q^A_R$ in the
mixture.  The samething is true on Bob's side. Without loss of
generality, we can restrict the analysis to protocols in which the
initial state on both sides is such a purification instead of a
mixture.  The attack will use the purification, but this is fine
because a cheater against a non purified protocol can himself create
such a purification in place of the mixture.  This is essentially the
same purification technique as in the original proof which ignored
superselection rules.  The difference is that here, in addition, we
make sure that all states in the initial superposition have the same
value for the conserved quantity.  Let $Q^A_R$ and $Q^B_R$ be the
fixed value for the conserved quantity on Alice's side and Bob's side,
respectively.  Because the protocol is known by both parties, we have
that $Q^A_R$ and $Q^B_R$ and thus $Q_R = Q^A_R + Q^B_R$ is known by
both parties.  The fact that bit commitment is impossible in the model
where $Q_R$ is fixed and known by both parties was proven
independently by Kitaev \cite{personal02} and the author.  The two
proofs are completely different. Here we present the author's proof.
We assume that the protocol is perfectly concealing and show that
Alice can swap bit~$0$ to bit~$1$.  The generalization to the inexact
case is not difficult.

After the commit phase, the overall state of the protocol associated
with an honest commitment of bit $w \in \{0,1\}$ can be written as
$\Psi(w) = \sum_{k = 0}^{Q_R} \sqrt{p_k(w)} \psi_k(w)$ where
\[
\psi_k(w) 
= |\Lambda|^{-1/2} \sum_{\lambda \in \Lambda}
\phi^A_{\lambda,k}(w) \otimes \phi^B_{\lambda,k}(w).
\]
and $\phi^A_{\lambda,k}(w)$ and $\phi^B_{\lambda,k}(w)$ are states
with $Q^A_R = k$ and $Q^B_R = Q_R - k$, respectively.  Both Alice and
Bob can measure $k$.  Because the protocol is concealing, we have that
$p_k(0) = p_k(1)$ for every $k$ because otherwise $k$ provides
information about $w$.  Let $\rho^B_k(w) = {\rm
Tr}_A(|\psi_k(w)\rangle\langle \psi_k(w) |)$ be the residual density
matrix on Bob's side associated with a given $k$ and $w$. Since Bob
can measure $k$ and then try to distinguish $\rho^B_k(0)$ and
$\rho^B_k(1)$ to obtain information about $w$, we have that
$\rho^B_k(0) = \rho^B_k(1)$, for every $k$.  The impossibility of bit
commitment applies to each $k$ individually. Therefore, conditioned on
$k$, Alice can execute a unitary transformation on her side which maps
every $\psi_k(0)$ exactly into $\psi_k(1)$. Because the different
$\psi_k(w)$ are in a superposition, the relative phase between these
states is important. This is not a problem because Alice has control
over this relative phase and she knows exactly the states $\psi_k(w)$,
including their relative phase. This concludes the proof.

\subsection*{Discussion}
We have shown that a charge conservation superselection rule imposes
no constraint on the allowed operations on every single qbit.
Therefore, even though neither Kitaev or Mayers were aware of this
fact at the time \cite{personal02}, the models that they used in their
respective impossibility proofs are actually valid even in the context
of such a superselection rule. In this way, we have addressed the
specific concern mentioned by Popescu \cite{popescu02}. It should be
possible to generalize this result to superselection rules that are
based on more general conservation laws.  An impossibility proof is no
more general than the model used. This is not at all a new
understanding for the author. In computational complexity, many models
were considered and only then Church and Turing proposed their general
thesis that all reasonable models are equivalent for the purpose of
computation.  The lesson that we learned here is that we should be
careful in quantum cryptography before we propose this kind of thesis
for quantum protocols.

On the experimental side, in view of our result, one may ask whether
or not it is experimentally difficult to swap a fermion mode and the
polarization of a photon.  This question is irrelevant if we are
interested in unconditional security, but nevertheless it is an
interesting question in itself.  We suspect that it is difficult and
this may have lead some people to the wrong conclusion that it is
fundamentally prohibited by a conservation law.  The required state
$(1/\sqrt n) \sum_{j = 0}^{n-1} |j\rangle^{(\vec y)}$ can in principle
be prepared with a non zero probability. One simply creates the state
$\otimes_{j=0}^{n-1} (1/\sqrt 2) (|0\rangle^{(y_j)} +
|1\rangle^{(y_j)})$ and then do a projection on the desired state.
The probability of success is small but more efficient algorithms can
most likely be designed.  In any case, in the context of unconditional
security, efficiency is not an issue. However, it is not clear whether
or not the individual states $(1/\sqrt 2) (|0\rangle^{(y_j)} +
|1\rangle^{(y_j)})$ are available in nature.  Moreover, a fully
functional quantum computer might not help because it might not be
able to manipulate systems that are restricted by the superselection
rule.

\end{document}